\documentclass[sigconf]{acmart}

\usepackage{enumitem}
\usepackage{listings}
\usepackage{tablefootnote}
\usepackage{xcolor}

\definecolor{codegreen}{rgb}{0,0.6,0}
\definecolor{codegray}{rgb}{0.5,0.5,0.5}
\definecolor{codepurple}{rgb}{0.58,0,0.82}
\lstdefinestyle{mystyle}{
    commentstyle=\color{codegreen},
    keywordstyle=\color{magenta},
    numberstyle=\tiny\color{codegray},
    stringstyle=\color{codepurple},
    basicstyle=\ttfamily\footnotesize,
    breakatwhitespace=false,        
    breaklines=true,                
    captionpos=b,                   
    keepspaces=true,                
    numbers=left,                   
    numbersep=5pt,                  
    showspaces=false,               
    showstringspaces=false,
    showtabs=false,                
    tabsize=2
}
\AtBeginDocument{%
  \providecommand\BibTeX{{%
    \normalfont B\kern-0.5em{\scshape i\kern-0.25em b}\kern-0.8em\TeX}}}

\acmSubmissionID{ICSE-WORKSHOP-LLM4Code-103}

\begin{document}

\title{PromptSet: A Programmer's Prompting Dataset}

\author{Kaiser Pister}
\email{kaiser@cs.wisc.edu}
\orcid{0009-0004-3144-8235}
\affiliation{%
  \institution{University of Wisconsin-Madison}
  \city{Madison}
  \country{USA}
}
\additionalaffiliation{
    \institution{Pister Labs}
}

\author{Dhruba Jyoti Paul}
\orcid{0009-0002-0817-6898}
\affiliation{%
  \institution{University of Wisconsin-Madison}
  \city{Madison}
  \country{USA}
}
\author{Patrick Brophy}
\orcid{0009-0004-0049-6736}
\affiliation{%
  \institution{University of Wisconsin-Madison}
  \city{Madison}
  \country{USA}
}
\author{Ishan Joshi}
\orcid{0009-0009-1298-2279}
\affiliation{%
  \institution{University of Wisconsin-Madison}
  \city{Madison}
  \country{USA}
}

\renewcommand{\shortauthors}{Pister, et al.}

% TODO: Revise and review
% TODO: Convert to links
\begin{abstract}
  The rise of capabilities expressed by large language models has been quickly followed by the integration of the same complex systems into application level logic. Algorithms, programs, systems, and companies are built around structured prompting to black box models where the majority of the design and implementation lies in capturing and quantifying the `agent mode'. The standard way to shape a closed language model is to prime it for a specific task with a tailored prompt, often initially handwritten by a human. The textual prompts co-evolve with the codebase, taking shape over the course of project life as artifacts which must be reviewed and maintained, just as the traditional code files might be. Unlike traditional code, we find that prompts do not receive effective static testing and linting to prevent runtime issues. In this work, we present a novel dataset called PromptSet, with more than 61,000 unique developer prompts used in open source Python programs. We perform analysis on this dataset and introduce the notion of a static linter for prompts. Released with this publication is a HuggingFace dataset and a Github repository to recreate collection and processing efforts, both under the name \texttt{pisterlabs/promptset}.
\end{abstract}
% effective first class treatment like the code objects they are
\copyrightyear{2024}
\acmYear{2024}
\setcopyright{acmlicensed}
\acmConference[LLM4Code '24]{2024 International Workshop on Large Language Models for Code}{April 20, 2024}{Lisbon, Portugal}
\acmBooktitle{2024 International Workshop on Large Language Models for Code (LLM4Code '24), April 20, 2024, Lisbon, Portugal}
\acmDOI{10.1145/3643795.3648395}
\acmISBN{979-8-4007-0579-3/24/04}




\received{25 January 2024}

\maketitle

\section{Introduction}
As large language models (LLMs) become more effective, more software is written to control and capture their potential. For any given task, a developer will have three common options to improve LLM performance: improve the foundational model, finetune a foundational model towards a specific task, or use a task tailored prompt. Major improvements across many benchmarks can be achieved by retraining the base model from scratch, however outside of large research labs, this is an intractable task \cite{openai2023gpt4, touvron2023llama, jiang2023mistral}. Finetuning can function as a cheaper alternative to retraining from scratch, but can require a modest sized dataset to achieve meaningful results. Additionally, it leads to catastrophic forgetting and complex model management when working with multiple tasks \cite{alpaca, lora}. Prompting can serve as a cheap, efficient way to control an LLM without significant investment in infrastructure or data \cite{2302.14691, NEURIPS2020_1457c0d6}. Since the introduction of few-shot learning, prompting has become a necessary feature of working with LLMs, and despite the proliferation of research into prompt engineering, in-context-learning and similar fields, there has been relatively little exploration of prompt management \cite{reynolds2021prompt, zhou2022large}. What's more, there exists no standard methodology for working with prompts, leading to incoherent collections of files, folders, JSON objects, and code strings inhibiting readability,  reusability, and maintainability \cite{jiang2020know}.

With LLMs moving from research into production, it is ever more important to treat the prompts which control business logic as first class citizens in the CI/CD pipelines of traditional software engineering. There are many works focusing on assessing and optimizing the effectiveness of a prompt, once it is in your system (either automatically or manually), but these rely on runtime computation against an LLM backbone and do not provide any static guarantees \cite{yang2023large, mekala2023echoprompt, zhao2021calibrate, lu2022fantastically}.

In this work, we propose static analysis passes to add to an existing CI/CD pipeline to preemptively detect non-traditional errors in prompts, such as misuse of variable formatting, typo detection, and input sanitization. In order to motivate these passes, we introduce a suite of extraction techniques to parse prompts from code files and a novel dataset, PromptSet, of more than 61,000 unique developer prompts used in open source Python programs. Finally, we are the first to our knowledge to discuss unit testing in the context of prompts as code.

\section{Background}
% Not talking about math
We abstract the language model to its functional purpose of decoding tokens based on an input sequence of tokens \cite{holtzman2023generativemodels}. The internal mechanisms are independent of the work presented here. For an in-depth understanding of the aforementioned architecture, we direct readers toward the existing literature on this subject. \cite{NIPS2017_3f5ee243, brown2020language, phuong2022formal}.

% Prompting tools background
Following the discovery of zero-shot and few-shot abilities of LLMs, a cottage industry of observation tools has developed in the market, allowing programmers to adequately monitor and experiment with open access language models \cite{opentel, langserve}. 
These tools give users the ability to detect anomalies in their agent systems, affording reliability in a design space that is notoriously unreliable. 
In parallel, quality is afforded by a large array of tools for optimizing the performance of individual prompts. 
Online courses for \texttt{prompt engineering} as a discipline have cropped up adjacent to Github repositories of "awesome-prompts" to improve a developer's manual prompting ability \cite{awesomeprompts, white2023prompt}. 
Although the topic is often criticised as a pseudo-science, common tricks and minor variations have demonstrably large effects on the quality of results \cite{anthlong}. 
Automatic optimization tools leverage an LLM as a self-optimizer across an established dataset to improve prompt quality \cite{yang2023large}.

In traditional software engineering reliability and quality are checked regularly through the use of continuous integration and delivery (CI/CD) systems.
CI/CD pipelines can manage many aspects of a code base, such as formatting, linting, and unit testing.
Regardless of the implementation, these systems are responsible for reducing the number of issues deployed to production systems. 
Formatting keeps code consistent, simplifying reviews between team members. 
Linting uses static analysis to keep a code base clean from common problems.
Unit testing guarantees run-time assumptions at a granular level.
As development teams scale to non-trivial sizes, CI/CD becomes one of the key factors in maintaining feature velocity \cite{instacd}.

Primarily due to the infancy of the field of prompting, we find the CI/CD ecosystem for prompting wanting. Formatting for prompts resides at the level of the developer's editor and is agnostic to the purpose of the prompt. Often a prompt exists alongside the rest of the code as a simple string.
There does not exist any form of linting for prompts, even bugs as simple as a mismatched variable interpolation won't be detected until run-time.

Functionality (integration) testing of prompts is quite popular, and there are many tools that support this behavior \cite{squidgy}.
We distinguish between the effectiveness of a prompt (its evaluation results on a dataset) and the characteristics of the prompt. For example, a unit test of a prompt might be an assertion that the prompt does not attempt to interpolate a string into an integer formatted slot (e.g. \texttt{"Num: \{:02d\}".format("x")}). 
A bug of this class could be detected by a static analysis pass, similar to any traditional type error.

\subsection{Taxonomies of Prompts}
Other works have looked to build a prescriptive taxonomy of effective prompting techniques \cite{white2023prompt, Hndler2023BalancingAA}. We leverage these works to categorize and classify our prompt dataset, but find that a large majority of our prompts do not fit neatly into one or more of these categories.
We take an unsupervised, descriptive approach with our classifications, and leave further category refinement to future work.

\subsection{Other Prompting Datasets}
At Mining Software Repositories '24, there was a code mining challenge addressing a similar task of understanding the purpose of prompts sent to DevGPT, a chat bot built for talking to code repositories on Github \cite{devgpt}. 
In contrast to the prompts we find in our dataset, DevGPT presents conversational text, similar to a user talking on ChatGPT. 
PromptSet represents programmatic prompts that often dictate application level logic. 
These prompts are not designed for chatting with a user, except occasionally by way of a programmed bot.

\section{Methodology}
\begin{figure*}[h]
  \centering
  \includegraphics[width=\linewidth]{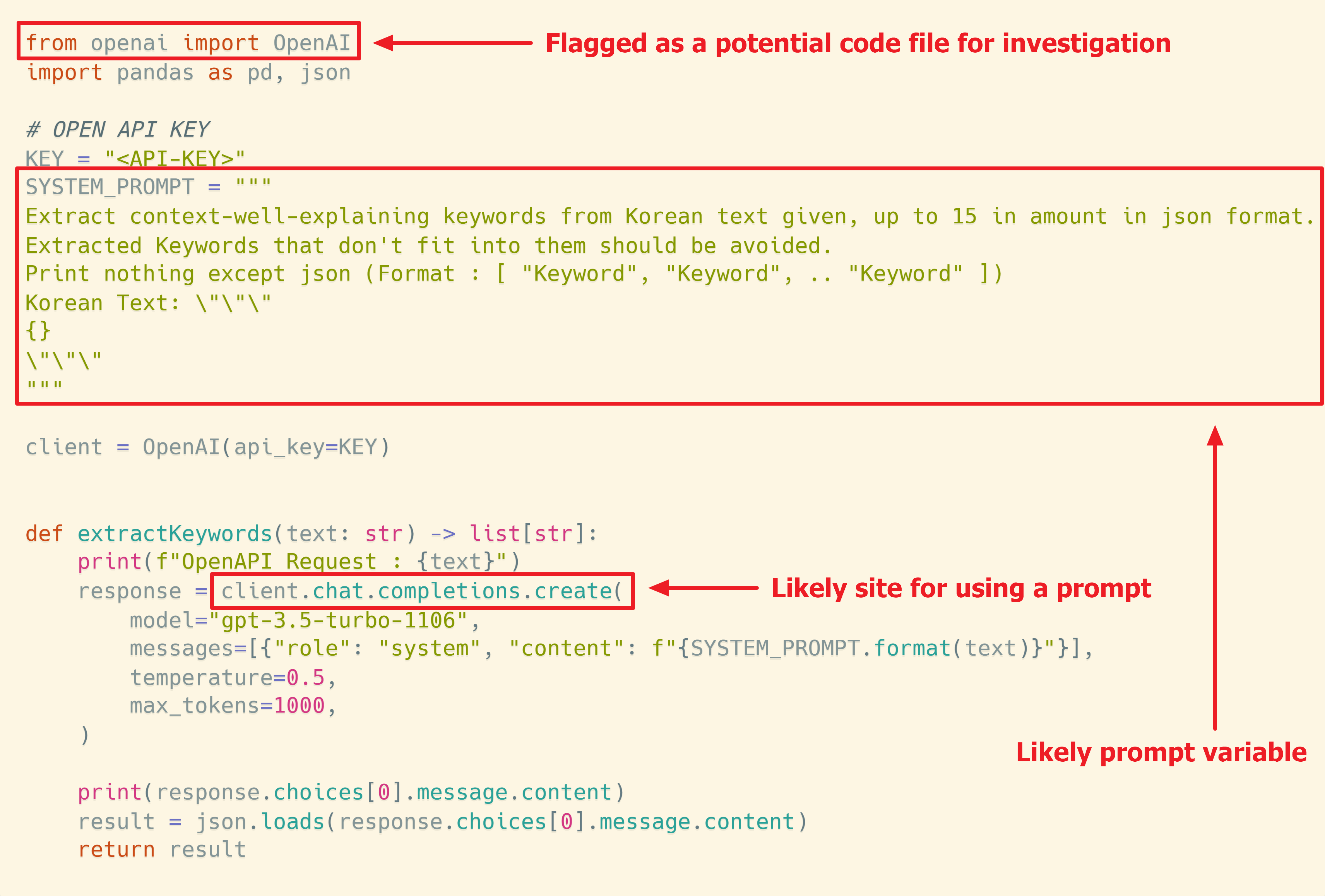}
  \caption{Example code file}
  \Description{Code file}
  \label{fig:snippet}
\end{figure*}

Towards the goal of developing meaningful static analysis passes, we first perform an ethnographic survey of how developers build applications with open LLM SDKs. The code survey is intended to validate our hypotheses and motivate the development of targeted linting rules or unit tests. We make PromptSet openly available to facilitate future static analysis from the community.

Our prompt dataset originates from open source code hosted on Github. We scrape Python code files which fulfill a simple criteria of using a language model library; see appendix A for the specific query string. We search for usage of the \texttt{openai}, \texttt{anthropic}, \texttt{cohere}, and \texttt{langchain} libraries, as these are regularly cited as the most popular tools in online resources. We ignore forked repositories to reduce duplication.
Notably, we exclude the \texttt{transformers} library which would include text prompts, but we find it to have too many false positives due to its other use cases.
Figure \ref{fig:snippet} shows an example code file which has been flagged for processing.

\subsection{Extracting Prompts}\label{extracting}
We utilize Tree-sitter for building abstract syntax tree (AST) representations of each scraped file and then query against the tree with specific patterns based on API design specifications and commonly observed behavior \cite{treesitter}.

We expect our dataset to be a subset of the entire set of prompts in these files as discussed in section \ref{limitations}.

\begin{enumerate}
    \item OpenAI \& Anthropic API calls: \\
        Both OpenAI (beta, v1) and Anthropic have similar API design specifications. 
        The SDKs expose \texttt{.create} functions on the \texttt{.completions} endpoint. 
        OpenAI additionally exposes a \texttt{.chat} variant, which is captured with the same pattern. 
        From the method calls, we extract specific argument values for positionally passed arguments, as well as the keyword arguments \texttt{prompt} and \texttt{messages}. 
        Common additional arguments are tracked and displayed in Table \ref{tab:args}.
    \item Cohere API calls: \\
        The Cohere SDK exposes multiple methods for handling text input. The two main entry points are \texttt{.chat} and \texttt{.summarize}. We pattern match on these calls, and follow the same methodology as above for extracting the arguments.
    \item LangChain PromptTemplate and Message classes: \\
        LangChain introduces multiple ways to create prompts, in particular a \texttt{PromptTemplate} class and a \texttt{Message}\\ (\texttt{HumanMessage}, \texttt{AIMessage}) class, which support completion and chat endpoints respectively. 
        These are base classes that developers can extend with their own functionality. 
        We search for constructor calls of classes which contain \texttt{*Prompt} or \texttt{*Message} in their name, and extract the arguments passed to the initialize function.
    \item LangChain Tools: \\
        LangChain exposes a Python decorator for automatically converting a function into an LLM tool using the function's docstring and type hints as the tool's description. Similarly, they allow for extending a \texttt{BaseTool} class. We match on \texttt{@tool} decorators and \texttt{BaseTool} super classes to detect these use cases.
    \item PromptTemplate.from\_file \\
        LangChain allows users to save prompts to standard text files, and load them at run time for processing. The \texttt{from\_file} function accomplishes this task, and we match directly on any use of $\texttt{*Template.from\_file}$.
    \item Prompt and Template variables \\
        From our observations, we see that the vast majority of variables used as prompts are named with the key phrase \texttt{prompt} or \texttt{template}. We flag any variable declarations matching this pattern, but note that this could be a noisy heuristic. In practice it proves to match the quality of the other extraction techniques.    
        \footnote{We found the \texttt{message} name was overloaded too often to be of much use.}
    \item "content" in dictionaries \\
        The basic structure for many chat messaging templates is using a Python dictionary with the keys \texttt{role} and \texttt{content}. The \texttt{role} entry typically contains "user", "system" or "assistant". The \texttt{content} entry contains the message that will be sent to LLM. We directly search for dictionaries with the specific key matching "content" to extract these prompts.
    \item DevGPT Conversation Prompts \\
        Finally, we add the prompts from the recent DevGPT MSR challenge to compare and contrast against PromptSet. These prompts are different in nature, and we keep them separate through our evaluation.
\end{enumerate}

\subsection{Post-Processing}
After extracting the key regions of the AST from each file, we use the \texttt{black} formatter to consistently format the inputs for better readability and deduplication \cite{black}.
Finally, we use tree-sitter again to extract strings, identifiers, and interpolations per prompt.

\subsection{Increasing Yield}
After performing a first iteration through these heuristics, we reviewed the extracted prompts as well as the files which failed to find any prompts and refined the extraction process. In particular, we added a naive $\beta$-substitution preprocessing step to replace constant expressions with their respective values throughout the file. As we process a file in search of prompt patterns, we track constant variable declarations by looking for static single assignment to variable names in the same file. Furthermore, we construct a simple string-evaluation language for processing string operations such as concatenation and interpolation across the static variable set. See listing \ref{lst:ex} for an example.

We note that this string interpreter is sound but not complete, and cannot process all Python string manipulations. Further exploration of partial execution models could be leveraged to increase the yield of prompts, but we leave this to future work.

\begin{lstlisting}[language=Python,label={lst:ex},caption=naive $\beta$-substitution]
import cohere  # file is flagged for processing

co = cohere.Client()
pre = "You are an agent working at the check-in desk." 
query = pre + " User said: {history}"
co.generate(query)  # flag `query`

# compute: query [pre:= "You are...", history: <free>]
# query := "You are an ... User said: PLACEHOLDER"
\end{lstlisting}

\subsection{Discarded Heuristics}
We consider a multi-line string heuristic but upon review found too many false positives given the correlation between the \texttt{streamlit} library, docstrings, and LLM development. We consider a sequence classifier to detect prompts but discard it for similar reasons.

\section{Results}
We describe PromptSet in two parts. First we provide a surface overview of the dataset, then we provide specific findings from our analysis. The prompt scraping and extraction was last run on January 10, 2024. By manually searching Github with our queries, we approximate that there are 153,000 code files which match our search criteria. Due to rate limiting, we are able to download 93,142 of these files, so we conclude that our dataset represents 60.7\% of the open-source API-based LLM-usage. The 93,142 code files come from 37,944 repositories.

\subsection{Dataset Overview}
Using the methodology described in section \ref{extracting}, we extract 118,862 total prompts from the scraped files, as seen in table \ref{tab:per_source}. The extracted prompts come from 37,112 of the code files (20,598 repositories). The remaining 56,030 files do not contain any prompts matching our extraction criteria, in part due to over scraping from Github and in part due to strict pattern matching. We perform a manual review of 200 code files which report no prompts found and manually tag 36 as false negatives, i.e. these files did contain prompts, but our extraction methodology did not find them. The majority of the true negatives are files that only use the semantic embedding functionality of these libraries or provide light wrappers around the library APIs. A quick check against the search terms on Github reveals that the ratio of extracted prompts matches roughly with the distribution between LangChain (most popular), OpenAI (popular), and Cohere (uncommon). Tool usage was introduced relatively recently, so the small count of tools is expected. 

\begin{table}
  \caption{Prompt Count per Source}
  \label{tab:per_source}
  \begin{tabular}{ccc}
    \toprule
    Library&Source&Count\\
    \midrule
    OpenAI/Anth.&completions.create & 12,420 \\
    Cohere&.chat & 260 \\
    LangChain&@tool & 1,425 \\
    LangChain&Template/Message class & 24,302 \\
    LangChain&from\_file & 21 \\
    All&Prompt/Template name & 94,897 \\
    All&Content Key in dictionary & 34,324\\
    DevGPT&Conversations&13,748\\
  \bottomrule
\end{tabular}
\end{table}

\begin{figure}[h]
  \centering
  \includegraphics[width=\linewidth]{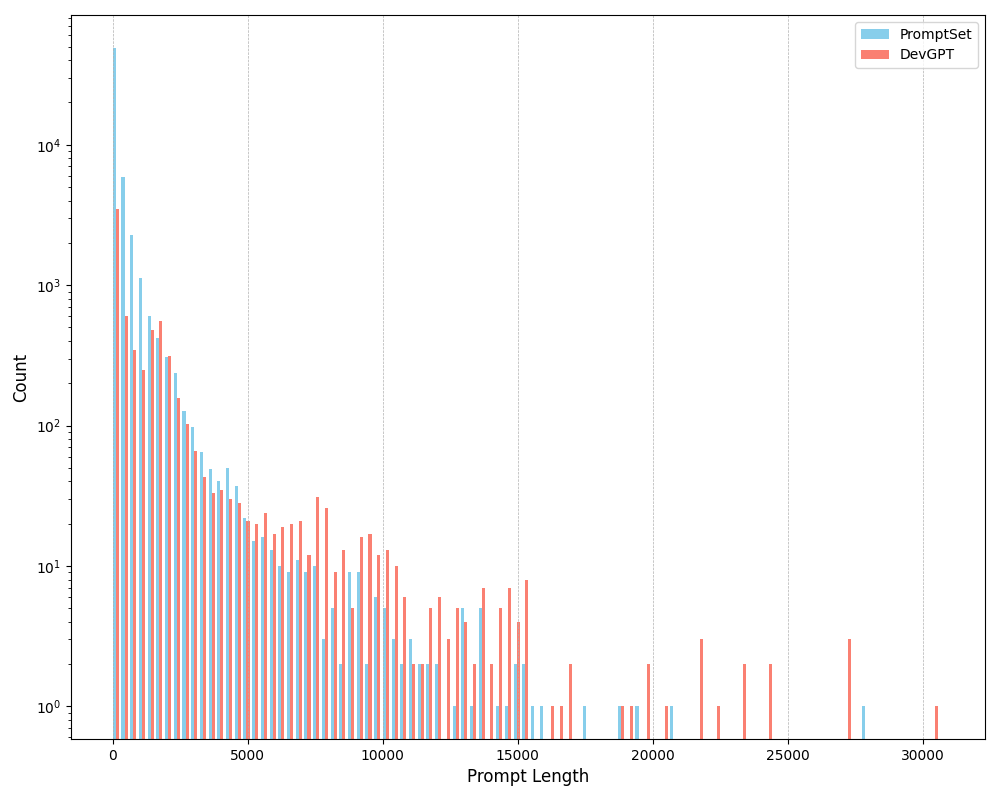}
  \caption{Distribution of prompt lengths in PromptSet.}
  \Description{A graph showing frequency of character lengths.}
  \label{fig:lengths}
\end{figure}

\begin{figure}[h]
  \centering
  \includegraphics[width=\linewidth]{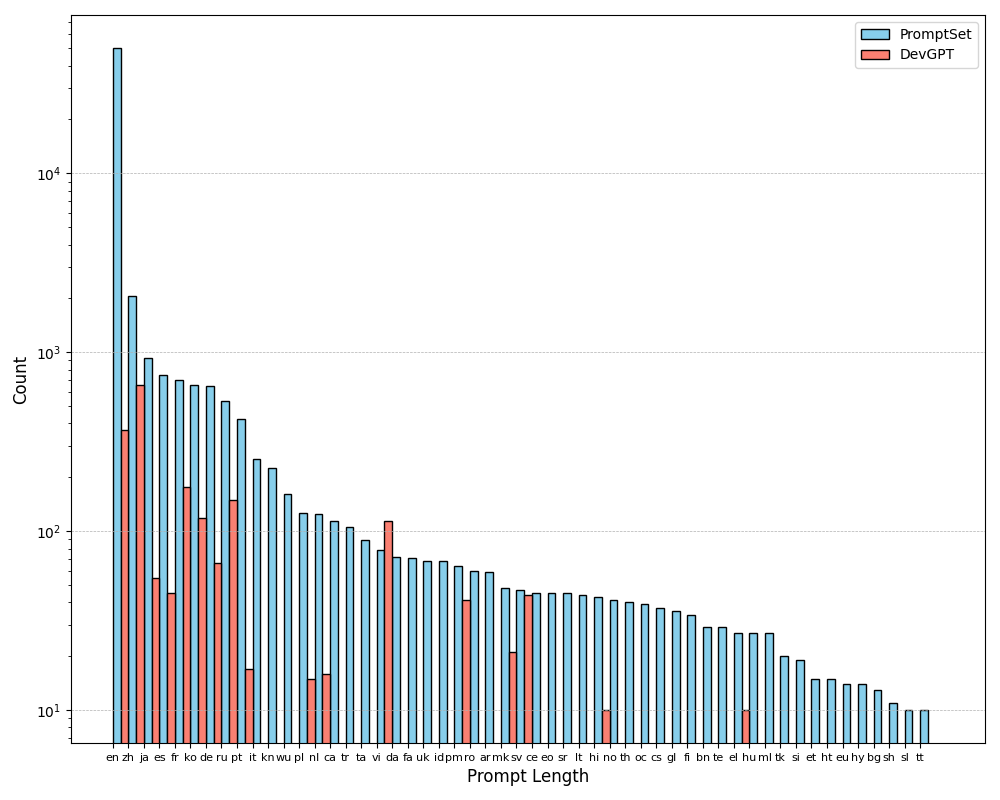}
  \caption{Distribution of languages in PromptSet.}
  \Description{A graph showing frequency of languages.}
  \label{fig:language}
\end{figure}

In order to perform per-prompt analysis, we join the prompts into a single set for testing. The results of deduplication are shown in table \ref{tab:unique_prompts}. We display length and language distributions of the dataset in figures \ref{fig:lengths} and \ref{fig:language}. Language is detected on prompts of length greater than ten characters using the \texttt{ftlangdetect} package \cite{joulin2016bag, joulin2016fasttext}. The majority of prompts are written in English, accounting for 84.1\% of the strings we extracted. The remaining 15.9\% fall between Mandarin, Japanese, Spanish, French, German, and Korean (below 1\% are not mentioned).

Briefly, we investigate the distributions of interpolations, and confirm that the most common variables are \texttt{chat}, \texttt{query}, \texttt{input} and similar placeholder input values for conversational AI. There was minimal representation of type-formatting statements in the dataset, (e.g. \texttt{"{ratio:.2f}"}), all of which were floating point formatting.

%TODO: remove this table?
\begin{table}
  \caption{Unique Prompts}
  \label{tab:unique_prompts}
  \begin{tabular}{ccccc}
    \toprule
    Set & Total Found & Unique & Length > 10 & Repositories\\
    \midrule
        PromptSet & 118,862 & 61,448 & 57,981 & 20,598 \\
        DevGPT & 13,748 & 13,236 & 13,053 & -  \\
  \bottomrule
\end{tabular}
\end{table}

\begin{table}
  \caption{LLM Call Arguments}
  \label{tab:args}
  \begin{tabular}{cccc}
    \toprule
    Parameter & Most Common\tablefootnote{Data reported from original November dataset} & 2nd & 3rd \\
    \midrule
        model & gpt-3.5-turbo & davinci-003 & gpt-4 \\
        temperature & 0 & 0.7 & 0.5 \\
        top\_p & 1 & 0.95 & 0.5  \\
        max\_tokens & 100 & 1024 & 1000 \\
  \bottomrule
\end{tabular}
\end{table}

\begin{figure}[h]
  \centering
  \includegraphics[width=\linewidth]{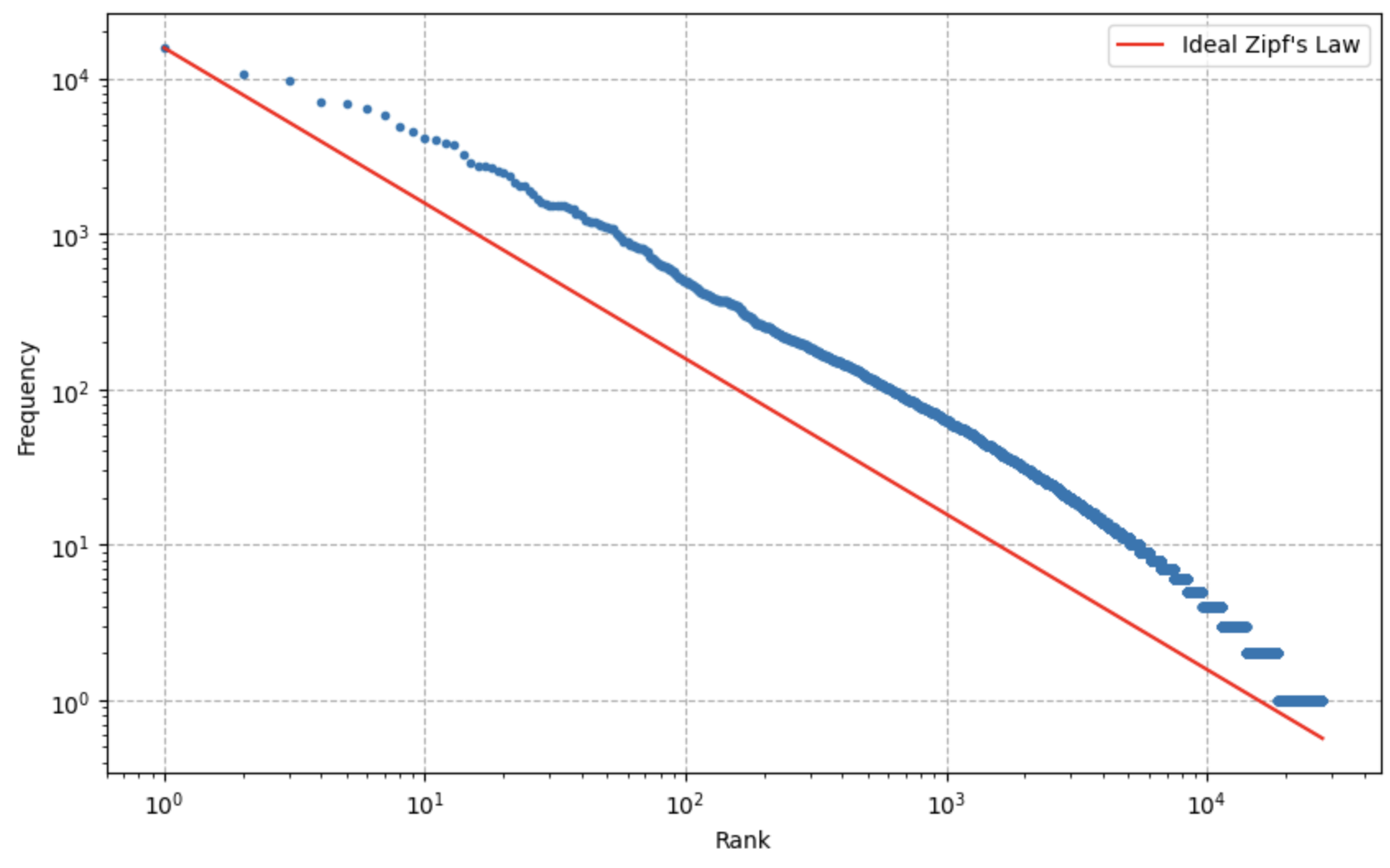}
  \caption{Zipf's law plotted on tokens from PromptSet.}
  \Description{A graph showing Zipf's law.}
  \label{fig:zipfs}
\end{figure}

Finally, we perform a Zipf's law analysis on the input tokens using the \texttt{cl100k\_base} tokenizer, as that supports the most common models used. From figure \ref{fig:zipfs}, we see that the mass is distributed above the ideal line, meaning there is a more even distribution across the token set than in traditional writing.

\subsection{Categorization and Clustering}
\begin{figure}[h]
  \centering
  \includegraphics[width=\linewidth]{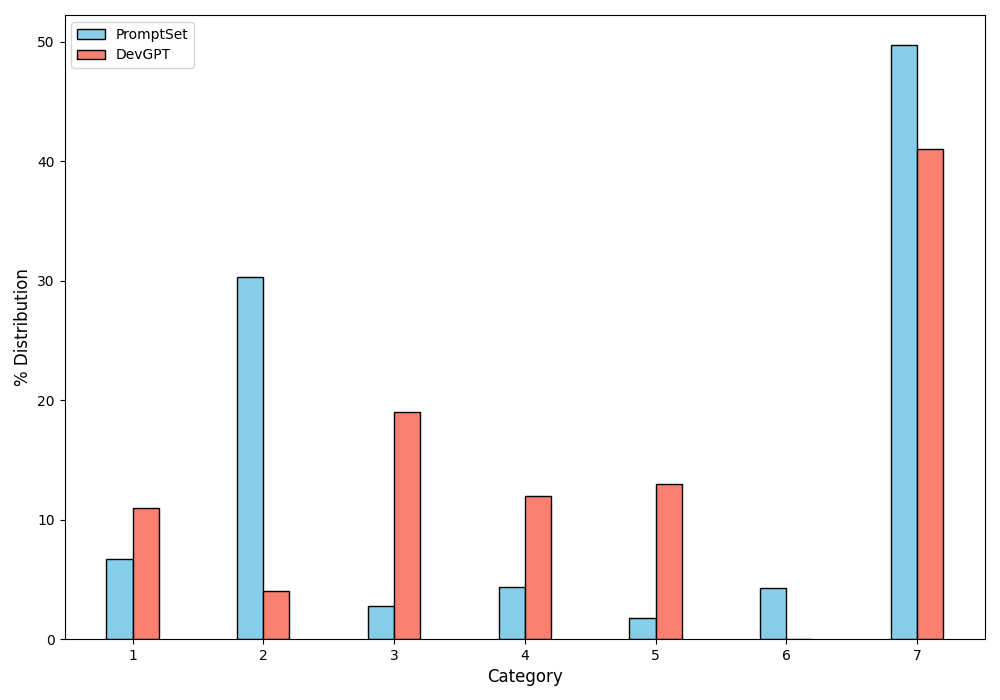}
  \caption{Categorization of PromptSet.}
  \Description{Plot showing the categorization of prompts.}
  \label{fig:classification}
\end{figure}

Using the dataset, we begin investigation in multiple directions to better understand the potential use cases of these prompts. To start, we follow the work of White et al. and categorize the prompts, using the six categories laid out in their work \cite{white2023prompt}. We craft a prompt based on their explanation of the categories and send 2,200 input prompts to the \texttt{gpt-4-preview-1106} model for prediction across PromptSet and DevGPT (2,000 for PromptSet, 200 for DevGPT). The results are shown in figure \ref{fig:classification} and an enumeration of the categories with extracted examples is provided in table \ref{tab:categories}.\footnote{Interestingly, only two of the prompts in the 2,000 PromptSet sampled prompts caused breaks in our prompt.}

\begin{table*}
\caption{Prompt Patterns Per White et al.}
\label{tab:categories}
\begin{tabular}{ccl}
        \toprule
        Pattern Name & Source & Example\tablefootnote{Prompts are abbreviated for space}  \\
        \midrule
        Input Semantics & jxnl & You are an expert at outputting json. \\ & & You always output valid JSON based on the pydantic schema given to you.  \\
        Output Customization & benczech212 & You are a wizard shop owner named \{ASSISTANT\_NAME\}. \\ & & Only talk on the behalf of \{ASSISTANT\_NAME\}. My name is \{USER\_NAME\}\\
        Error Identification & & \\
        Prompt Improvement & Kaastor & Given the following conversation and a follow up question, rephrase the follow up question \\ &&to be a standalone question.  \\
        Interaction & offtian & You are playing the '20 Questions' game with another player. Your role is to answer 'Yes' or \\ && 'No' to questions based on a given concept or object.  \\
        Context Control & nachollorca & Your task is to answer a question given some context given here, delimited by triple backticks: \\
        \bottomrule
\end{tabular}
\end{table*}

These results show that there is some alignment between the prescribed "good prompting techniques" and the prompting techniques we find in the wild, however there are still large discrepancies.
Many users are enacting the category-2, "Output Customization-Persona", beginning their prompts with "Act as a...", to elicit a specific type of response.
The distribution of DevGPT prompts is more evenly spread than the distributions of PromptSet which favors categories 1 and 2 (Input and Output). We believe this to be the case because developers require strict control of the input and output to their systems. Only once those are under control can they leverage categories 3-6. On the other hand, in a conversation there are no such restrictions.

Many of the prompts fail to fall into any of the categories, and we suspect this is due to a few contributing factors. First, some of the prompts are partial prompts which might no have a clear category without more context. Second, since the prompts were labeled with an LLM with a meta-prompt that did not undergo any optimization, there is the possibility for error. Third, the taxonomy proposed by White et al. is now ten months old, and the study of prompting has progressed much since its release. It is possible that a new category could emerge, though we do not see a clear trend in the dataset at this point.

\begin{figure}[h]
  \centering
  \includegraphics[width=\linewidth]{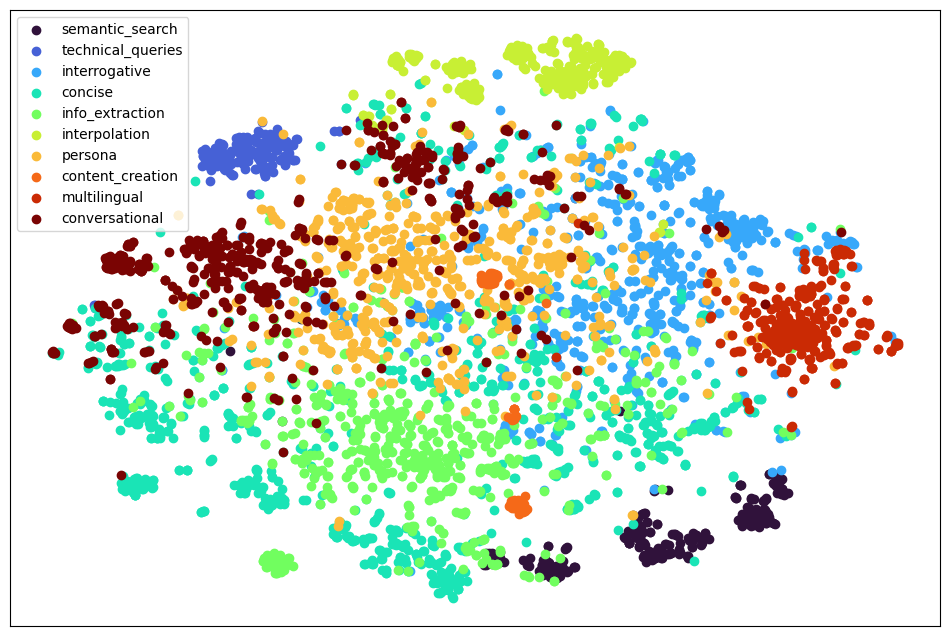}
  \caption{t-SNE of PromptSet.}
  \Description{Plot showing the t-SNE of prompt embeddings.}
  \label{fig:tsne}
\end{figure}

In order to perform our own classification, we create a clustering nearest neighbor plot with the semantic meaning of each prompt. To start, we embed each prompt using the \texttt{all-MiniLM-L6-v2} model from the \texttt{sentence-transformers} library \cite{senttran}. We fit a t-SNE with 10 clusters to the embedding outputs and show the results in figure \ref{fig:tsne}. Manual inspection of the clusters shows many similarities, and we assign labeled names based on the most common patterns we see.

\subsection{Technique Propagation}
Next we investigate the propagation of research techniques into PromptSet in table \ref{tab:technique}. We use a few heuristics to derive usage of some of the most popular techniques, such as chain-of-thought, few shot prompting, and special tokens \cite{wei2023chainofthought, besta2023graph}. For each of these techniques, we assign a few representative strings to filter on through each of the splits. Chain-of-thought, for example, matches with "step-by-step", "step by step", "let(')s think", and "thought(s):". We expect this to be an underestimate on usage. The \texttt{doc} column represents prompts mentioning "documents" which have a high prevalence in recent product developments.

\begin{table*}
\caption{Research Technique Detection}
\label{tab:technique}
\begin{tabular}{cccccccccccc}
        \toprule
        Set & Total & concise & Few Shot & doc & CoT & Code Block & Instruction Block & Scratchpad & Tool use & Special Tokens \\
        \midrule
        PromptSet & 57953 
& 176 (0.3) & 1008 (1.7) & 1939 (3.3) & 1095 (1.9) & 1696 (2.9) & 927 (1.6) & 170 (0.3) & 168 (0.3) & 178 (0.3) \\
        DevSet & 13053 
& 2 (0.0) & 88 (0.7) & 1015 (7.8) & 105 (0.8) & 730 (5.6) & 120 (0.9) & 0 (0.0) & 319 (2.4) & 16 (0.1) \\
        \bottomrule
\end{tabular}
\end{table*}

\subsection{Error Investigation}
To perform an error investigation, we first look for typos in the prompts. Prompts tend to be natural language requests, typed by hand and as we have observed, can be riddled with typos. We use the \texttt{cspell} package on each prompt and find close to 28,000 spelling errors across the unique English prompts of PromptSet \cite{cspell}. Many of the spelling errors stem from proper nouns and code related terminology, so we remove capitalized mistakes and words with underscores, reducing the error count to 16,989. To further improve the accuracy two authors independently manually tag 200 of these errors, and find a true positive rate between 33\%-40\%. If we extrapolate with this rate, there would be a typo in approximately 1 of 8 prompts.

These mistakes are not limited to junior developers. In writing this work, we found typos in our own prompts as well as the academic papers that we have reviewed \cite{staab2023memorization}.

Finally, we develop a simple white space detection lint pass, which checks for trailing and leading white space in prompts. The official documentation from OpenAI discusses that trailing white spaces in prompts can lead to poor tokenization which causes a degradation in performance of the model. While a simple \texttt{.strip()} can resolve the issue, there is no guarantee that this stripping happens on the server side. Thus white space detection is a perfect problem for a linter to solve. The results in table \ref{tab:whitespace} show that a large portion of the prompts we find do indeed have trailing white spaces, such as newlines, tabs and spaces.

\begin{table}
  \caption{Leading \& Trailing Whitespace Detection}
  \label{tab:whitespace}
  \begin{tabular}{ccccc}
        \toprule
        Set & Total & Trailing (\%) & Leading (\%) & All (\%) \\
        \midrule
        PromptSet & 58,814 & 17,668 (30.0) & 9,723 (16.5) & 19,681 (33.5) \\
        DevGPT & 13,399 & 2,081 (15.5) & 235 (1.8) & 2,256 (16.8) \\
        \bottomrule
\end{tabular}
\end{table}
\section{Discussion}
In this work we introduce a novel dataset with the purpose of better understanding how developers are interacting with the newfound power of integrating LLMs into their applications. PromptSet indeed displays a diversity of ideas and we acknowledge that even this only represents a fraction of the programming prompt usage that exists.

\subsection{Limitations}\label{limitations}
As mentioned in the results, we expect to have approximately 60\% public coverage of the libraries we mention, however there are other libraries we did not consider scraping and many closed source repositories that are obviously out of reach. Additionally, we restrict our set to Python, but there is an active JavaScript development community focused on LLM development as well. Beyond hitting API limits to fully search Github, the extraction techniques proposed do not find the full set of prompt strings used in the files mentioned, nor do they extract prompts from adjacent files in the same system (for example prompts that were imported from another file). This means that while PromptSet contains a large number of diverse prompts, it might not reflect the true distribution and characteristics of prompts. Consequently, our dataset is not exhaustive and may not include all relevant data points, a factor that must be considered when interpreting the findings, as it could lead to gaps that affect the overall results.
\subsection{Future Work}

Despite the potential benefit to readability, we cannot quantify that typos or prompt mistakes are bad for every possible task \cite{min2022rethinking}. Instead we posit that in the majority of the cases, a developer would like to actively make a choice in handling the likely mistakes of typos.

A good unit test for a prompt might test that the prompt follows the project guidelines on appropriate wording, or asserts that the prompt does not allow for injection into a non-data section of the prompt \cite{huang2023catastrophic}. Perhaps asserting that all prompts include the proper persona within a repository would be a helpful test for some developers. With unit testing, the power lies in the flexibility to tailor the test to each individual task and prompt.

The goal for PromptSet is to put forward a tool to parse prompts from files so that downstream applications can easily perform proper prompt management. We hope to establish a conversation about appropriate prompt hygiene so that the open source community can develop strong tooling for improving the CI/CD pipeline for prompts. In this work we propose a few surface level lint passes, such as typo detection, white space trimming and type annotation matching, however the possibilities go far beyond these simple tests.

\bibliographystyle{ACM-Reference-Format}
\bibliography{base}


\begin{thebibliography}{38}

%%% the Web address.

\ifx \showCODEN    \undefined \def \showCODEN     #1{\unskip}     \fi
\ifx \showDOI      \undefined \def \showDOI       #1{#1}\fi
\ifx \showISBNx    \undefined \def \showISBNx     #1{\unskip}     \fi
\ifx \showISBNxiii \undefined \def \showISBNxiii  #1{\unskip}     \fi
\ifx \showISSN     \undefined \def \showISSN      #1{\unskip}     \fi
\ifx \showLCCN     \undefined \def \showLCCN      #1{\unskip}     \fi
\ifx \shownote     \undefined \def \shownote      #1{#1}          \fi
\ifx \showarticletitle \undefined \def \showarticletitle #1{#1}   \fi
\ifx \showURL      \undefined \def \showURL       {\relax}        \fi
% invisible to TeX
\providecommand\bibfield[2]{#2}
\providecommand\bibinfo[2]{#2}
\providecommand\natexlab[1]{#1}
\providecommand\showeprint[2][]{arXiv:#2}

\bibitem[AI(2023)]%
        {langserve}
\bibfield{author}{\bibinfo{person}{LangChain AI}.} \bibinfo{year}{2023}\natexlab{}.
\newblock \bibinfo{title}{LangServe}.
\newblock
\newblock
\urldef\tempurl%
\url{https://github.com/langchain-ai/langserve}
\showURL{%
\tempurl}


\bibitem[Anthropic(2023)]%
        {anthlong}
\bibfield{author}{\bibinfo{person}{Anthropic}.} \bibinfo{year}{2023}\natexlab{}.
\newblock \bibinfo{title}{Claude 2.1 Prompting}.
\newblock
\newblock
\urldef\tempurl%
\url{https://www.anthropic.com/index/claude-2-1-prompting}
\showURL{%
\tempurl}


\bibitem[Besta et~al\mbox{.}(2023)]%
        {besta2023graph}
\bibfield{author}{\bibinfo{person}{Maciej Besta}, \bibinfo{person}{Nils Blach}, \bibinfo{person}{Ales Kubicek}, \bibinfo{person}{Robert Gerstenberger}, \bibinfo{person}{Lukas Gianinazzi}, \bibinfo{person}{Joanna Gajda}, \bibinfo{person}{Tomasz Lehmann}, \bibinfo{person}{Michal Podstawski}, \bibinfo{person}{Hubert Niewiadomski}, \bibinfo{person}{Piotr Nyczyk}, {and} \bibinfo{person}{Torsten Hoefler}.} \bibinfo{year}{2023}\natexlab{}.
\newblock \bibinfo{title}{Graph of Thoughts: Solving Elaborate Problems with Large Language Models}.
\newblock
\newblock
\showeprint[arxiv]{2308.09687}~[cs.CL]


\bibitem[Brown et~al\mbox{.}(2020a)]%
        {NEURIPS2020_1457c0d6}
\bibfield{author}{\bibinfo{person}{Tom Brown}, \bibinfo{person}{Benjamin Mann}, \bibinfo{person}{Nick Ryder}, \bibinfo{person}{Melanie Subbiah}, \bibinfo{person}{Jared~D Kaplan}, \bibinfo{person}{Prafulla Dhariwal}, \bibinfo{person}{Arvind Neelakantan}, \bibinfo{person}{Pranav Shyam}, \bibinfo{person}{Girish Sastry}, \bibinfo{person}{Amanda Askell}, \bibinfo{person}{Sandhini Agarwal}, \bibinfo{person}{Ariel Herbert-Voss}, \bibinfo{person}{Gretchen Krueger}, \bibinfo{person}{Tom Henighan}, \bibinfo{person}{Rewon Child}, \bibinfo{person}{Aditya Ramesh}, \bibinfo{person}{Daniel Ziegler}, \bibinfo{person}{Jeffrey Wu}, \bibinfo{person}{Clemens Winter}, \bibinfo{person}{Chris Hesse}, \bibinfo{person}{Mark Chen}, \bibinfo{person}{Eric Sigler}, \bibinfo{person}{Mateusz Litwin}, \bibinfo{person}{Scott Gray}, \bibinfo{person}{Benjamin Chess}, \bibinfo{person}{Jack Clark}, \bibinfo{person}{Christopher Berner}, \bibinfo{person}{Sam McCandlish}, \bibinfo{person}{Alec Radford}, \bibinfo{person}{Ilya Sutskever}, {and}
  \bibinfo{person}{Dario Amodei}.} \bibinfo{year}{2020}\natexlab{a}.
\newblock \showarticletitle{Language Models are Few-Shot Learners}. In \bibinfo{booktitle}{\emph{Advances in Neural Information Processing Systems}}, \bibfield{editor}{\bibinfo{person}{H.~Larochelle}, \bibinfo{person}{M.~Ranzato}, \bibinfo{person}{R.~Hadsell}, \bibinfo{person}{M.F. Balcan}, {and} \bibinfo{person}{H.~Lin}} (Eds.), Vol.~\bibinfo{volume}{33}. \bibinfo{publisher}{Curran Associates, Inc.}, \bibinfo{pages}{1877--1901}.
\newblock
\urldef\tempurl%
\url{https://proceedings.neurips.cc/paper_files/paper/2020/file/1457c0d6bfcb4967418bfb8ac142f64a-Paper.pdf}
\showURL{%
\tempurl}


\bibitem[Brown et~al\mbox{.}(2020b)]%
        {brown2020language}
\bibfield{author}{\bibinfo{person}{Tom~B. Brown}, \bibinfo{person}{Benjamin Mann}, \bibinfo{person}{Nick Ryder}, \bibinfo{person}{Melanie Subbiah}, \bibinfo{person}{Jared Kaplan}, \bibinfo{person}{Prafulla Dhariwal}, \bibinfo{person}{Arvind Neelakantan}, \bibinfo{person}{Pranav Shyam}, \bibinfo{person}{Girish Sastry}, \bibinfo{person}{Amanda Askell}, \bibinfo{person}{Sandhini Agarwal}, \bibinfo{person}{Ariel Herbert-Voss}, \bibinfo{person}{Gretchen Krueger}, \bibinfo{person}{Tom Henighan}, \bibinfo{person}{Rewon Child}, \bibinfo{person}{Aditya Ramesh}, \bibinfo{person}{Daniel~M. Ziegler}, \bibinfo{person}{Jeffrey Wu}, \bibinfo{person}{Clemens Winter}, \bibinfo{person}{Christopher Hesse}, \bibinfo{person}{Mark Chen}, \bibinfo{person}{Eric Sigler}, \bibinfo{person}{Mateusz Litwin}, \bibinfo{person}{Scott Gray}, \bibinfo{person}{Benjamin Chess}, \bibinfo{person}{Jack Clark}, \bibinfo{person}{Christopher Berner}, \bibinfo{person}{Sam McCandlish}, \bibinfo{person}{Alec Radford}, \bibinfo{person}{Ilya Sutskever},
  {and} \bibinfo{person}{Dario Amodei}.} \bibinfo{year}{2020}\natexlab{b}.
\newblock \bibinfo{title}{Language Models are Few-Shot Learners}.
\newblock
\newblock
\showeprint[arxiv]{2005.14165}~[cs.CL]


\bibitem[Corralm(2023)]%
        {awesomeprompts}
\bibfield{author}{\bibinfo{person}{Miguel Corralm}.} \bibinfo{year}{2023}\natexlab{}.
\newblock \bibinfo{title}{Awesome Prompting}.
\newblock
\newblock
\urldef\tempurl%
\url{https://github.com/corralm/awesome-prompting}
\showURL{%
\tempurl}


\bibitem[CSpell(2023)]%
        {cspell}
\bibfield{author}{\bibinfo{person}{CSpell}.} \bibinfo{year}{2023}\natexlab{}.
\newblock \bibinfo{title}{CSpell}.
\newblock
\newblock
\urldef\tempurl%
\url{https://www.npmjs.com/package/cspell}
\showURL{%
\tempurl}


\bibitem[Foundation(2023)]%
        {black}
\bibfield{author}{\bibinfo{person}{Python~Software Foundation}.} \bibinfo{year}{2023}\natexlab{}.
\newblock \bibinfo{title}{Black}.
\newblock
\newblock
\urldef\tempurl%
\url{https://github.com/psf/black}
\showURL{%
\tempurl}


\bibitem[H{\"a}ndler(2023)]%
        {Hndler2023BalancingAA}
\bibfield{author}{\bibinfo{person}{Thorsten H{\"a}ndler}.} \bibinfo{year}{2023}\natexlab{}.
\newblock \showarticletitle{Balancing Autonomy and Alignment: A Multi-Dimensional Taxonomy for Autonomous LLM-powered Multi-Agent Architectures}.
\newblock \bibinfo{journal}{\emph{ArXiv}}  \bibinfo{volume}{abs/2310.03659} (\bibinfo{year}{2023}).
\newblock
\urldef\tempurl%
\url{https://api.semanticscholar.org/CorpusID:263671545}
\showURL{%
\tempurl}


\bibitem[Holtzman et~al\mbox{.}(2023)]%
        {holtzman2023generativemodels}
\bibfield{author}{\bibinfo{person}{Ari Holtzman}, \bibinfo{person}{Peter West}, {and} \bibinfo{person}{Luke Zettlemoyer}.} \bibinfo{year}{2023}\natexlab{}.
\newblock \showarticletitle{Generative Models as a Complex Systems Science: How can we make sense of large language model behavior?}
\newblock \bibinfo{journal}{\emph{preprint}} (\bibinfo{year}{2023}).
\newblock


\bibitem[Hu et~al\mbox{.}(2022)]%
        {lora}
\bibfield{author}{\bibinfo{person}{Edward~J Hu}, \bibinfo{person}{Yelong Shen}, \bibinfo{person}{Phillip Wallis}, \bibinfo{person}{Zeyuan Allen-Zhu}, \bibinfo{person}{Yuanzhi Li}, \bibinfo{person}{Shean Wang}, \bibinfo{person}{Lu Wang}, {and} \bibinfo{person}{Weizhu Chen}.} \bibinfo{year}{2022}\natexlab{}.
\newblock \showarticletitle{Lo{RA}: Low-Rank Adaptation of Large Language Models}. In \bibinfo{booktitle}{\emph{International Conference on Learning Representations}}.
\newblock
\urldef\tempurl%
\url{https://openreview.net/forum?id=nZeVKeeFYf9}
\showURL{%
\tempurl}


\bibitem[Huang et~al\mbox{.}(2023)]%
        {huang2023catastrophic}
\bibfield{author}{\bibinfo{person}{Yangsibo Huang}, \bibinfo{person}{Samyak Gupta}, \bibinfo{person}{Mengzhou Xia}, \bibinfo{person}{Kai Li}, {and} \bibinfo{person}{Danqi Chen}.} \bibinfo{year}{2023}\natexlab{}.
\newblock \showarticletitle{Catastrophic Jailbreak of Open-source LLMs via Exploiting Generation}.
\newblock \bibinfo{journal}{\emph{arXiv preprint arXiv:2310.06987}} (\bibinfo{year}{2023}).
\newblock


\bibitem[Instagram(2016)]%
        {instacd}
\bibfield{author}{\bibinfo{person}{Instagram}.} \bibinfo{year}{2016}\natexlab{}.
\newblock \bibinfo{title}{Continuous Deployment at Instagram}.
\newblock
\newblock
\urldef\tempurl%
\url{https://instagram-engineering.com/continuous-deployment-at-instagram-1e18548f01d1}
\showURL{%
\tempurl}


\bibitem[Jiang et~al\mbox{.}(2023)]%
        {jiang2023mistral}
\bibfield{author}{\bibinfo{person}{Albert~Q. Jiang}, \bibinfo{person}{Alexandre Sablayrolles}, \bibinfo{person}{Arthur Mensch}, \bibinfo{person}{Chris Bamford}, \bibinfo{person}{Devendra~Singh Chaplot}, \bibinfo{person}{Diego de~las Casas}, \bibinfo{person}{Florian Bressand}, \bibinfo{person}{Gianna Lengyel}, \bibinfo{person}{Guillaume Lample}, \bibinfo{person}{Lucile Saulnier}, \bibinfo{person}{Lélio~Renard Lavaud}, \bibinfo{person}{Marie-Anne Lachaux}, \bibinfo{person}{Pierre Stock}, \bibinfo{person}{Teven~Le Scao}, \bibinfo{person}{Thibaut Lavril}, \bibinfo{person}{Thomas Wang}, \bibinfo{person}{Timothée Lacroix}, {and} \bibinfo{person}{William~El Sayed}.} \bibinfo{year}{2023}\natexlab{}.
\newblock \bibinfo{title}{Mistral 7B}.
\newblock
\newblock
\showeprint[arxiv]{2310.06825}~[cs.CL]


\bibitem[Jiang et~al\mbox{.}(2020)]%
        {jiang2020know}
\bibfield{author}{\bibinfo{person}{Zhengbao Jiang}, \bibinfo{person}{Frank~F. Xu}, \bibinfo{person}{Jun Araki}, {and} \bibinfo{person}{Graham Neubig}.} \bibinfo{year}{2020}\natexlab{}.
\newblock \bibinfo{title}{How Can We Know What Language Models Know?}
\newblock
\newblock
\showeprint[arxiv]{1911.12543}~[cs.CL]


\bibitem[Joulin et~al\mbox{.}(2016b)]%
        {joulin2016fasttext}
\bibfield{author}{\bibinfo{person}{Armand Joulin}, \bibinfo{person}{Edouard Grave}, \bibinfo{person}{Piotr Bojanowski}, \bibinfo{person}{Matthijs Douze}, \bibinfo{person}{H{\'e}rve J{\'e}gou}, {and} \bibinfo{person}{Tomas Mikolov}.} \bibinfo{year}{2016}\natexlab{b}.
\newblock \showarticletitle{FastText.zip: Compressing text classification models}.
\newblock \bibinfo{journal}{\emph{arXiv preprint arXiv:1612.03651}} (\bibinfo{year}{2016}).
\newblock


\bibitem[Joulin et~al\mbox{.}(2016a)]%
        {joulin2016bag}
\bibfield{author}{\bibinfo{person}{Armand Joulin}, \bibinfo{person}{Edouard Grave}, \bibinfo{person}{Piotr Bojanowski}, {and} \bibinfo{person}{Tomas Mikolov}.} \bibinfo{year}{2016}\natexlab{a}.
\newblock \showarticletitle{Bag of Tricks for Efficient Text Classification}.
\newblock \bibinfo{journal}{\emph{arXiv preprint arXiv:1607.01759}} (\bibinfo{year}{2016}).
\newblock


\bibitem[Lu et~al\mbox{.}(2022)]%
        {lu2022fantastically}
\bibfield{author}{\bibinfo{person}{Yao Lu}, \bibinfo{person}{Max Bartolo}, \bibinfo{person}{Alastair Moore}, \bibinfo{person}{Sebastian Riedel}, {and} \bibinfo{person}{Pontus Stenetorp}.} \bibinfo{year}{2022}\natexlab{}.
\newblock \bibinfo{title}{Fantastically Ordered Prompts and Where to Find Them: Overcoming Few-Shot Prompt Order Sensitivity}.
\newblock
\newblock
\showeprint[arxiv]{2104.08786}~[cs.CL]


\bibitem[Mekala et~al\mbox{.}(2023)]%
        {mekala2023echoprompt}
\bibfield{author}{\bibinfo{person}{Rajasekhar~Reddy Mekala}, \bibinfo{person}{Yasaman Razeghi}, {and} \bibinfo{person}{Sameer Singh}.} \bibinfo{year}{2023}\natexlab{}.
\newblock \bibinfo{title}{EchoPrompt: Instructing the Model to Rephrase Queries for Improved In-context Learning}.
\newblock
\newblock
\showeprint[arxiv]{2309.10687}~[cs.CL]


\bibitem[Min et~al\mbox{.}(2022)]%
        {min2022rethinking}
\bibfield{author}{\bibinfo{person}{Sewon Min}, \bibinfo{person}{Xinxi Lyu}, \bibinfo{person}{Ari Holtzman}, \bibinfo{person}{Mikel Artetxe}, \bibinfo{person}{Mike Lewis}, \bibinfo{person}{Hannaneh Hajishirzi}, {and} \bibinfo{person}{Luke Zettlemoyer}.} \bibinfo{year}{2022}\natexlab{}.
\newblock \bibinfo{title}{Rethinking the Role of Demonstrations: What Makes In-Context Learning Work?}
\newblock
\newblock
\showeprint[arxiv]{2202.12837}~[cs.CL]


\bibitem[OpenAI(2023)]%
        {openai2023gpt4}
\bibfield{author}{\bibinfo{person}{OpenAI}.} \bibinfo{year}{2023}\natexlab{}.
\newblock \bibinfo{title}{GPT-4 Technical Report}.
\newblock
\newblock
\showeprint[arxiv]{2303.08774}~[cs.CL]


\bibitem[Phuong and Hutter(2022)]%
        {phuong2022formal}
\bibfield{author}{\bibinfo{person}{Mary Phuong} {and} \bibinfo{person}{Marcus Hutter}.} \bibinfo{year}{2022}\natexlab{}.
\newblock \bibinfo{title}{Formal Algorithms for Transformers}.
\newblock
\newblock
\showeprint[arxiv]{2207.09238}~[cs.LG]


\bibitem[Reimers and Gurevych(2019)]%
        {senttran}
\bibfield{author}{\bibinfo{person}{Nils Reimers} {and} \bibinfo{person}{Iryna Gurevych}.} \bibinfo{year}{2019}\natexlab{}.
\newblock \showarticletitle{Sentence-BERT: Sentence Embeddings using Siamese BERT-Networks}.
\newblock \bibinfo{journal}{\emph{CoRR}}  \bibinfo{volume}{abs/1908.10084} (\bibinfo{year}{2019}).
\newblock
\showeprint[arXiv]{1908.10084}
\urldef\tempurl%
\url{http://arxiv.org/abs/1908.10084}
\showURL{%
\tempurl}


\bibitem[Reynolds and McDonell(2021)]%
        {reynolds2021prompt}
\bibfield{author}{\bibinfo{person}{Laria Reynolds} {and} \bibinfo{person}{Kyle McDonell}.} \bibinfo{year}{2021}\natexlab{}.
\newblock \bibinfo{title}{Prompt Programming for Large Language Models: Beyond the Few-Shot Paradigm}.
\newblock
\newblock
\showeprint[arxiv]{2102.07350}~[cs.CL]


\bibitem[SquidgyAI(2023)]%
        {squidgy}
\bibfield{author}{\bibinfo{person}{SquidgyAI}.} \bibinfo{year}{2023}\natexlab{}.
\newblock \bibinfo{title}{Squidgy Testy}.
\newblock
\newblock
\urldef\tempurl%
\url{https://github.com/squidgyai/squidgy-testy}
\showURL{%
\tempurl}


\bibitem[Staab et~al\mbox{.}(2023)]%
        {staab2023memorization}
\bibfield{author}{\bibinfo{person}{Robin Staab}, \bibinfo{person}{Mark Vero}, \bibinfo{person}{Mislav Balunović}, {and} \bibinfo{person}{Martin Vechev}.} \bibinfo{year}{2023}\natexlab{}.
\newblock \bibinfo{title}{Beyond Memorization: Violating Privacy Via Inference with Large Language Models}.
\newblock
\newblock
\showeprint[arxiv]{2310.07298}~[cs.AI]


\bibitem[Taori et~al\mbox{.}(2023)]%
        {alpaca}
\bibfield{author}{\bibinfo{person}{Rohan Taori}, \bibinfo{person}{Ishaan Gulrajani}, \bibinfo{person}{Tianyi Zhang}, \bibinfo{person}{Yann Dubois}, \bibinfo{person}{Xuechen Li}, \bibinfo{person}{Carlos Guestrin}, \bibinfo{person}{Percy Liang}, {and} \bibinfo{person}{Tatsunori~B. Hashimoto}.} \bibinfo{year}{2023}\natexlab{}.
\newblock \bibinfo{title}{Stanford Alpaca: An Instruction-following LLaMA model}.
\newblock \bibinfo{howpublished}{\url{https://github.com/tatsu-lab/stanford_alpaca}}.
\newblock


\bibitem[Touvron et~al\mbox{.}(2023)]%
        {touvron2023llama}
\bibfield{author}{\bibinfo{person}{Hugo Touvron}, \bibinfo{person}{Louis Martin}, \bibinfo{person}{Kevin Stone}, \bibinfo{person}{Peter Albert}, \bibinfo{person}{Amjad Almahairi}, \bibinfo{person}{Yasmine Babaei}, \bibinfo{person}{Nikolay Bashlykov}, \bibinfo{person}{Soumya Batra}, \bibinfo{person}{Prajjwal Bhargava}, \bibinfo{person}{Shruti Bhosale}, \bibinfo{person}{Dan Bikel}, \bibinfo{person}{Lukas Blecher}, \bibinfo{person}{Cristian~Canton Ferrer}, \bibinfo{person}{Moya Chen}, \bibinfo{person}{Guillem Cucurull}, \bibinfo{person}{David Esiobu}, \bibinfo{person}{Jude Fernandes}, \bibinfo{person}{Jeremy Fu}, \bibinfo{person}{Wenyin Fu}, \bibinfo{person}{Brian Fuller}, \bibinfo{person}{Cynthia Gao}, \bibinfo{person}{Vedanuj Goswami}, \bibinfo{person}{Naman Goyal}, \bibinfo{person}{Anthony Hartshorn}, \bibinfo{person}{Saghar Hosseini}, \bibinfo{person}{Rui Hou}, \bibinfo{person}{Hakan Inan}, \bibinfo{person}{Marcin Kardas}, \bibinfo{person}{Viktor Kerkez}, \bibinfo{person}{Madian Khabsa},
  \bibinfo{person}{Isabel Kloumann}, \bibinfo{person}{Artem Korenev}, \bibinfo{person}{Punit~Singh Koura}, \bibinfo{person}{Marie-Anne Lachaux}, \bibinfo{person}{Thibaut Lavril}, \bibinfo{person}{Jenya Lee}, \bibinfo{person}{Diana Liskovich}, \bibinfo{person}{Yinghai Lu}, \bibinfo{person}{Yuning Mao}, \bibinfo{person}{Xavier Martinet}, \bibinfo{person}{Todor Mihaylov}, \bibinfo{person}{Pushkar Mishra}, \bibinfo{person}{Igor Molybog}, \bibinfo{person}{Yixin Nie}, \bibinfo{person}{Andrew Poulton}, \bibinfo{person}{Jeremy Reizenstein}, \bibinfo{person}{Rashi Rungta}, \bibinfo{person}{Kalyan Saladi}, \bibinfo{person}{Alan Schelten}, \bibinfo{person}{Ruan Silva}, \bibinfo{person}{Eric~Michael Smith}, \bibinfo{person}{Ranjan Subramanian}, \bibinfo{person}{Xiaoqing~Ellen Tan}, \bibinfo{person}{Binh Tang}, \bibinfo{person}{Ross Taylor}, \bibinfo{person}{Adina Williams}, \bibinfo{person}{Jian~Xiang Kuan}, \bibinfo{person}{Puxin Xu}, \bibinfo{person}{Zheng Yan}, \bibinfo{person}{Iliyan Zarov}, \bibinfo{person}{Yuchen
  Zhang}, \bibinfo{person}{Angela Fan}, \bibinfo{person}{Melanie Kambadur}, \bibinfo{person}{Sharan Narang}, \bibinfo{person}{Aurelien Rodriguez}, \bibinfo{person}{Robert Stojnic}, \bibinfo{person}{Sergey Edunov}, {and} \bibinfo{person}{Thomas Scialom}.} \bibinfo{year}{2023}\natexlab{}.
\newblock \bibinfo{title}{Llama 2: Open Foundation and Fine-Tuned Chat Models}.
\newblock
\newblock
\showeprint[arxiv]{2307.09288}~[cs.CL]


\bibitem[Traceloop(2023)]%
        {opentel}
\bibfield{author}{\bibinfo{person}{Traceloop}.} \bibinfo{year}{2023}\natexlab{}.
\newblock \bibinfo{title}{OpenTelemetry}.
\newblock
\newblock
\urldef\tempurl%
\url{https://www.traceloop.com/blog/diy-observability-for-llm-with-opentelemetry}
\showURL{%
\tempurl}


\bibitem[tree sitter({[n.\,d.]})]%
        {treesitter}
\bibfield{author}{\bibinfo{person}{tree sitter}.} \bibinfo{year}{[n.\,d.]}\natexlab{}.
\newblock \bibinfo{title}{Tree-sitter}.
\newblock
\newblock
\urldef\tempurl%
\url{https://tree-sitter.github.io/tree-sitter}
\showURL{%
\tempurl}


\bibitem[Vaswani et~al\mbox{.}(2017)]%
        {NIPS2017_3f5ee243}
\bibfield{author}{\bibinfo{person}{Ashish Vaswani}, \bibinfo{person}{Noam Shazeer}, \bibinfo{person}{Niki Parmar}, \bibinfo{person}{Jakob Uszkoreit}, \bibinfo{person}{Llion Jones}, \bibinfo{person}{Aidan~N Gomez}, \bibinfo{person}{\L~ukasz Kaiser}, {and} \bibinfo{person}{Illia Polosukhin}.} \bibinfo{year}{2017}\natexlab{}.
\newblock \showarticletitle{Attention is All you Need}. In \bibinfo{booktitle}{\emph{Advances in Neural Information Processing Systems}}, \bibfield{editor}{\bibinfo{person}{I.~Guyon}, \bibinfo{person}{U.~Von Luxburg}, \bibinfo{person}{S.~Bengio}, \bibinfo{person}{H.~Wallach}, \bibinfo{person}{R.~Fergus}, \bibinfo{person}{S.~Vishwanathan}, {and} \bibinfo{person}{R.~Garnett}} (Eds.), Vol.~\bibinfo{volume}{30}. \bibinfo{publisher}{Curran Associates, Inc.}
\newblock
\urldef\tempurl%
\url{https://proceedings.neurips.cc/paper_files/paper/2017/file/3f5ee243547dee91fbd053c1c4a845aa-Paper.pdf}
\showURL{%
\tempurl}


\bibitem[Wei et~al\mbox{.}(2023)]%
        {wei2023chainofthought}
\bibfield{author}{\bibinfo{person}{Jason Wei}, \bibinfo{person}{Xuezhi Wang}, \bibinfo{person}{Dale Schuurmans}, \bibinfo{person}{Maarten Bosma}, \bibinfo{person}{Brian Ichter}, \bibinfo{person}{Fei Xia}, \bibinfo{person}{Ed Chi}, \bibinfo{person}{Quoc Le}, {and} \bibinfo{person}{Denny Zhou}.} \bibinfo{year}{2023}\natexlab{}.
\newblock \bibinfo{title}{Chain-of-Thought Prompting Elicits Reasoning in Large Language Models}.
\newblock
\newblock
\showeprint[arxiv]{2201.11903}~[cs.CL]


\bibitem[White et~al\mbox{.}(2023)]%
        {white2023prompt}
\bibfield{author}{\bibinfo{person}{Jules White}, \bibinfo{person}{Quchen Fu}, \bibinfo{person}{Sam Hays}, \bibinfo{person}{Michael Sandborn}, \bibinfo{person}{Carlos Olea}, \bibinfo{person}{Henry Gilbert}, \bibinfo{person}{Ashraf Elnashar}, \bibinfo{person}{Jesse Spencer-Smith}, {and} \bibinfo{person}{Douglas~C. Schmidt}.} \bibinfo{year}{2023}\natexlab{}.
\newblock \bibinfo{title}{A Prompt Pattern Catalog to Enhance Prompt Engineering with ChatGPT}.
\newblock
\newblock
\showeprint[arxiv]{2302.11382}~[cs.SE]


\bibitem[Xiao et~al\mbox{.}(2024)]%
        {devgpt}
\bibfield{author}{\bibinfo{person}{Tao Xiao}, \bibinfo{person}{Christoph Treude}, \bibinfo{person}{Hideaki Hata}, {and} \bibinfo{person}{Kenichi Matsumoto}.} \bibinfo{year}{2024}\natexlab{}.
\newblock \showarticletitle{DevGPT: Studying Developer-ChatGPT Conversations}. In \bibinfo{booktitle}{\emph{Proceedings of the International Conference on Mining Software Repositories (MSR 2024)}}.
\newblock


\bibitem[Yang et~al\mbox{.}(2023)]%
        {yang2023large}
\bibfield{author}{\bibinfo{person}{Chengrun Yang}, \bibinfo{person}{Xuezhi Wang}, \bibinfo{person}{Yifeng Lu}, \bibinfo{person}{Hanxiao Liu}, \bibinfo{person}{Quoc~V. Le}, \bibinfo{person}{Denny Zhou}, {and} \bibinfo{person}{Xinyun Chen}.} \bibinfo{year}{2023}\natexlab{}.
\newblock \bibinfo{title}{Large Language Models as Optimizers}.
\newblock
\newblock
\showeprint[arxiv]{2309.03409}~[cs.LG]


\bibitem[Ye et~al\mbox{.}(2023)]%
        {2302.14691}
\bibfield{author}{\bibinfo{person}{Seonghyeon Ye}, \bibinfo{person}{Hyeonbin Hwang}, \bibinfo{person}{Sohee Yang}, \bibinfo{person}{Hyeongu Yun}, \bibinfo{person}{Yireun Kim}, {and} \bibinfo{person}{Minjoon Seo}.} \bibinfo{year}{2023}\natexlab{}.
\newblock \bibinfo{title}{In-Context Instruction Learning}.
\newblock
\newblock
\showeprint{arXiv:2302.14691}


\bibitem[Zhao et~al\mbox{.}(2021)]%
        {zhao2021calibrate}
\bibfield{author}{\bibinfo{person}{Tony~Z. Zhao}, \bibinfo{person}{Eric Wallace}, \bibinfo{person}{Shi Feng}, \bibinfo{person}{Dan Klein}, {and} \bibinfo{person}{Sameer Singh}.} \bibinfo{year}{2021}\natexlab{}.
\newblock \bibinfo{title}{Calibrate Before Use: Improving Few-Shot Performance of Language Models}.
\newblock
\newblock
\showeprint[arxiv]{2102.09690}~[cs.CL]


\bibitem[Zhou et~al\mbox{.}(2022)]%
        {zhou2022large}
\bibfield{author}{\bibinfo{person}{Yongchao Zhou}, \bibinfo{person}{Andrei~Ioan Muresanu}, \bibinfo{person}{Ziwen Han}, \bibinfo{person}{Keiran Paster}, \bibinfo{person}{Silviu Pitis}, \bibinfo{person}{Harris Chan}, {and} \bibinfo{person}{Jimmy Ba}.} \bibinfo{year}{2022}\natexlab{}.
\newblock \showarticletitle{Large Language Models Are Human-Level Prompt Engineers}.
\newblock  (\bibinfo{year}{2022}).
\newblock
\showeprint[arxiv]{2211.01910}~[cs.LG]


\end{thebibliography}

\appendix

\section{Github Scraping Code}
\begin{lstlisting}[language=Python]
for lib in ["openai", 
    "anthropic", 
    "cohere", 
    "langchain"]:
    goto(f"https://github.com/search?" +
        "q=%22from+{lib}%22+OR+" +
        "%22import+{lib}%22+" + 
        "language%3Apython&type=code")
\end{lstlisting}

\end{document}